# STUDY OF THE DYNAMICS OF THE CORE OF A2218

**Priyamvada Natarajan**

*Institute of Astronomy, Madingley Road, Cambridge CB3 OHA, U.K.*

**Abstract.** We report the results of an attempt to study the velocity structure of the core of the cluster A2218 in order to probe its dynamical state. The aim is to understand if the core is indeed virialized? relaxed? or in hydrodynamic equilibrium? We use cluster lensing data in conjunction with the optical galaxy data to solve in detail for the nature of galaxy orbits and the velocity anisotropy in the core. We present our preliminary results below.

## 1. Introduction

Studying the velocity structure of the cores of clusters of galaxies promises to provide new insights into the physics of the formation of clusters. In order to obtain physical solutions it is crucial to know the mass distribution a priori. We are proposing a new method wherein we use the mass profile as inferred, independent of the dynamics, from gravitational lensing.

## 2. Previous Work

The velocity structure of galaxies in the Coma cluster have been studied in detail by The & White, Merritt and Kent & Gunn and several other groups. These analyses involved using the observed galaxy positions and velocities to constrain the distribution of total mass and thereby find consistent solutions for the velocity anisotropy. Many different mass models and hence diverse velocity profiles were found to be consistent with the data despite the restrictive and simplifying assumption that light traces mass. The data was insufficient to rule out any velocity anisotropy.

Similar studies of the cluster A2670 have also produced ambiguous results. For example, even in the faint photometric and spectroscopic survey by Sharples, Ellis & Gray, there were not many galaxies far out enough from the cluster center to plot a detailed velocity histogram to get an indication of the degree of departure from the isotropic Gaussian profile. Both extremely anisotropic models and nearly isotropic ones were indistinguishable in terms of goodness of fit with respect to the available data. In these clusters and others that have been studied, most of the uncertainty in the determination of the anisotropy arises from ignorance of the underlying mass profile in conjunction with the fact that the line-of-sight velocity dispersion is also poorly determined.

## 3. The Formalism

Since our new technique involves a priori knowledge of the mass profile obtained independently from gravitational lensing we eliminate the largest source of uncertainty. Schematically, from the observational galaxy data, we fit to get a surface number density profile $\Sigma_g(r)$ and use the Abel integral inversion to obtain $\rho_g(r)$. The two key assumptions made in the analysis below are :

(i) spherical symmetry and that

(ii) $M_{tot}(r)$ derived from lensing is indeed the accurate profile for the underlying mass distribution.

We model the cluster using the Jeans hydrodynamic equation for a collisionless system in equilibrium,

$$\frac{d}{dr}(\rho_g\ \sigma_r^2) + \frac{2\beta(r)\rho_g(r)\ \sigma_r^2}{r} = \frac{-\ G\ M_{tot}(r)\rho_g(r)}{r^2}$$

where

- $\rho_g(r)$ - galaxy density profile,
- $\sigma_r^2(r)$ - radial velocity dispersion of the galaxies,
- $\beta(r)$ - the velocity anisotropy defined as follows, $\beta(r) = \left(1 - \frac{\sigma_t^2}{\sigma_r^2}\right)$
- $\sigma_t^2(r)$ - tangential component of the velocity dispersion and
- $M_{tot}(r)$ is the distribution of the total mass from lensing.

In addition to the Jeans equation, we have the equation that defines the observed line-of-sight velocity dispersion,

$$\frac{1}{2}\Sigma_g(R)\ \sigma_{los}^2(R) = \int_R^\infty \frac{r\rho_g(r)\sigma_r^2(r)dr}{\sqrt{(r^2-R^2)}} - R^2 \int_R^\infty \frac{\beta(r)\sigma_r^2(r)\rho_g(r)dr}{r\sqrt{(r^2-R^2)}}$$

We solve the two integro-differential equations for $\sigma_r^2(r)$ and $\beta(r)$.

## 4. Application to A2218

We apply this technique to the Abell cluster A2218, at a redshift $z = 0.175$ with a mean measured velocity dispersion $\sigma_{mean} \sim 1370 kms^{-1}$. We use the photometric data and velocity data from the catalog published by Le Borgne et. al. A2218 is a cD cluster with a very peaked mass distribution with a compact core and hence a large number of gravitationally distorted arcs and arclets. The mass model (bimodal) for this cluster was constructed using ground data and refined by Kneib et. al.(1995a) using HST data. Redshifts of two of the arcs have been spectroscopically confirmed by Pello et. al. and therefore the mass model is tightly constrained.

Figure 1.  The velocity anisotropy parameter

### 4.1. Mass modeling from lensing

The mass profile for the cluster was constructed from the strong lensing data (arcs, arclets and multiple images) primarily from the HST image (Kneib, J. P., et. al. 1995a) has the following form,

$$M_{tot}(r) = M_o(\frac{r}{r_o} - tan^{-1}\frac{r}{r_o})$$

where $M_o = 3.149 \times 10^{13} M\odot$ and $r_o = 45h^{-1} kpc$ is the lensing core radius. This profile is valid within $r \leq 500h^{-1}\ kpc$ from the cluster center.

### 4.2. Galaxy density profile

The surface density of galaxies in A2218 was fitted to a Hubble law profile with a core radius $r_g = 250h^{-1}\ kpc$,

$$\rho_g(r) = \frac{\rho_{og}}{(1 + \frac{r^2}{r_g^2})^{\frac{3}{2}}}$$

### 5. Preliminary Results

Using the total mass profile constructed as above, we solve the equations to obtain the following solutions for $\sigma_r(r)$ and $\sigma_t(r)$ (the radial and tangential velocity dispersion profiles respectively) the velocity anisotropy parameter $\beta(r)$.

We find that the orbits in the core are primarily **transverse**, consistent with the picture of a virialized core. Both the radial velocity dispersion and the transverse component of the velocity dispersion fall within the inner 600 $kpc$ but $\sigma_r$ declines more rapidly and then tends to flatten off. At larger radii we expect the orbits to be mostly radial due to infall but the data at present are insufficient to probe the transition region. The run of the velocity anisotropy parameter with radius is very sensitive to details of the observed line-of-sight velocity dispersion profile which is also inadequately measured at present.

## 6. Conclusions and Future Work

- This technique offers a better understanding of the dynamics of cluster cores.

- With more optical galaxy data can put stronger constraints on the velocity structure in A2218.

- Using the reconstructed mass profile from weak lensing in conjunction with the above can extend the the distance out to which the orbits can be probed.

**Acknowledgments.** PN acknowledges useful discussions with Jean-Paul Kneib, Martin Rees and Simon White.